# Enhanced Super-Radiance in Epsilon-Near-Zero Plasmonic Channels

Romain Fleury, and Andrea Alù[*]

Dept. of Electrical & Computer Engineering, The University of Texas at Austin

[*]alu@mail.utexas.edu

*We describe the possibility of drastically boosting the spontaneous emission of a collection of two-level quantum emitters by embedding them in an epsilon-near-zero (ENZ) environment, consisting of a plasmonic waveguide operated at cut-off. This phenomenon relies on the combination of Purcell enhancement and Dicke superradiance effects, exploiting the large and uniform local density of states in ENZ channels, which is shown to significantly extend the spatial extent of these quantum effects. We envision exciting applications in optical sensing, molecular detection and low-threshold lasing.*

Enhancing the spontaneous emission of quantum emitters has been the subject of active research for over 50 years. Before the work of Purcell [1-2], spontaneous emission was considered an intrinsic property of atoms or molecules. He was probably the first to realize that the spontaneous decay rate largely depends on the environment and can be enhanced when coupled to an external resonance. A variety of modern applications rely on this phenomenon, including fluorescence microscopy [3], DNA sequencing [4], low-threshold lasing [5], single molecule detection [6] and surface enhanced Raman spectroscopy [7]. These applications generally rely on strongly coupling the emitters with resonant cavities [5], and more recently with plasmonic layers, nanoparticles or nanoshells [8-11]. In all these scenarios, the emission enhancement is always associated with an inherently limited extent of the available volume, requiring accurate positioning of the quantum emitters to take full advantage of the location where the density of states is maximized. Obviously, this directly limits the practicality of this concept and its applicability to a large number of quantum emitters collectively radiating.

Another possibility to achieve enhanced emission is based on the collective effects arising when $N$ quantum sources are placed sufficiently close to each other, known as super-radiance. Cooperative effects in spontaneous emission were first investigated by Dicke [12], who considered the collective emission of identical atoms confined in a deeply subwavelength volume. He showed that the resulting radiation intensity was not proportional to $N$, as one would expect for classically independent sources, but rather to $N^2$, with a radiation decay time no longer constant but proportional to $N^{-1}$. Also in this case, the effect works only provided that the sources are packed very closely, in order to maximize the number of collaborating emitters while maintaining a very small total electrical size [13]. This inherent limitation on the extent of available volume also here practically hinders the applicability of this effect.

Recent advances in metamaterials and plasmonics may provide a way to overcome the limitations of both approaches and combine them to achieve giant super-radiant effects from a collection of emitters [14]. Particularly interesting in this context is the physics of epsilon-near-zero (ENZ) metamaterials, which enable a counterintuitive transmission effect, named supercoupling, based on anomalous impedance matching between waveguides with largely different cross-sections [15-21]. Supercoupling has been theoretically suggested and experimentally realized at optical frequencies using the natural dispersion of narrow plasmonic channels operated at their cut-off wavelength, whose optical properties were shown to be equivalent to realizing an effective near-zero permittivity metamaterial [16-20]. In this unusual operation, the plasmonic channel, when excited from the outside, supports extraordinary optical transmission with very *large* field enhancements and, due to the inherently large phase velocity in ENZ materials, *uniform* field amplitude all along the channel, independent of the total channel length [17]. This concept was experimentally verified in [22] using cathodoluminescence measurements, and its use has been recently proposed to drastically boost optical nonlinearities in an array configuration [23].

Using Lorentz reciprocity, the largely enhanced and uniform field distribution in the ENZ channel implies that the Green's function and local density of states is similarly boosted *independent* of the location inside the channel. In this condition, Purcell effects would be largely enhanced over a scale potentially significantly larger than in conventional resonant systems, not requiring the perfect source positioning usually limiting other methods. This concept has been indeed proposed to show that several classical dipoles oscillating in phase and *randomly* placed along such plasmonic channel would produce a steady-state radiated field drastically enhanced at its ENZ operation [24-25].

In the classical treatment used in [24], the dipoles were assumed to be oscillating coherently and in phase, but it is interesting to analyze the quantum effects associated with this anomalous emission. In the following, we explore the problem from the quantum perspective, showing not only that randomly placed quantum emitters may strongly radiate in the channel independent of their position and relative distance, but also that Dicke super-radiance effects may further boost this effect and lead to giant coherent emission near the ENZ operation. Both effects are strongly enhanced by the fact that the effective emission wavelength in the channel is very large at the ENZ frequency, providing an interesting way to increase Purcell and cooperation lengths of a resonant plasmonic system. This results in an overall emission significantly larger than what envisioned in [24] using purely classical arguments, unveiling uniquely novel phenomena associated with combining classical ENZ metamaterials with quantum systems.

We first investigate the spontaneous emission properties of a single fluorescent atom placed inside a plasmonic rectangular waveguide with small cross-section, highlighting the exotic properties associated with Purcell enhancement in a narrow channel operating at the ENZ frequency. We then include a second atom and solve the problem for the coupled system, analyzing the magnitude and spatial range of Dicke cooperative effects in the ENZ environment compared to vacuum. Then, we extend these concepts to randomly placed emitters in large numbers, achieving a dramatic increase in radiated intensity based on the combination of largely boosted Purcell and Dicke effects.

Consider the geometry of Fig.1, depicting a narrow plasmonic waveguide of height $a_{ch}=40$ nm, width $b_{ch}=200$ nm and length $l_{ch}=1\,\mu$m, used to connect two semi-infinite rectangular waveguides of width $b=400$ nm and height $a=350\,nm \gg a_{ch}$. All waveguides are filled with vacuum with permittivity $\varepsilon_0$, and the metallic walls are made of silver modeled with a Drude permittivity $\varepsilon_{Ag}=\varepsilon_0(\varepsilon_\infty - f_p^2 / f(f+i\Gamma))$, $f_p=2175$ THz, $\Gamma=4.35$ THz and $\varepsilon_\infty=5$ [26]. Using transmission-line (TL) theory [19] and the effective index method [20], Fig.1 shows the line impedances and guided propagation constants of the small channel and the outer waveguides as a function of frequency. At a frequency $f_{ENZ} = 400\,\text{THZ}$, corresponding to the cut-off frequency of the narrow channel, the guided wave number in the channel $\beta_{ch} \simeq 0$, and impedance matching occurs at the two abruptions, due to the anomalous near-zero value of effective permittivity that compensates the large cross-sectional mismatch, enabling electromagnetic tunneling through the channel. At this specific frequency, supercoupling can occur independently of the channel length, with almost infinite phase velocity and uniform field distributions [17].

Consider now the spontaneous emission from an initially excited two-level quantum emitter embedded in the narrow channel. The decay rate of a two-level quantum system with emission frequency $\omega$ may be written as a function of the partial local density of states (LDOS) $\rho_d(\vec{r},\omega)$ at its position $\vec{r}$, assuming dipolar emission. The LDOS can be evaluated classically by calculating the dyadic Green's function $\vec{n}_d . \ddot{\vec{G}}(\vec{r},\vec{r}',\omega).\vec{n}_d$ at the dipole's position $\vec{r}' = \vec{r}$ [27-28]. For simplicity, we assume here that the transitions of the quantum system have fixed dipole axis $\vec{n}_d$ along $\vec{y}$ and we neglect recombination or quenching effects due to the vicinity to the metallic surface.

We use TL theory to calculate $\ddot{\vec{G}}(\vec{r},\vec{r},\omega)$ and compare the spontaneous emission rate $\gamma$ to its value $\gamma_0$ in vacuum. We have also used COMSOL full-wave simulations to validate our results, taking into account the abruption reactances that are neglected in our TL model, obtaining consistent results. In Fig. 2a, we show the calculated gain $\gamma/\gamma_0$ for a quantum emitter located at $z$ on the channel axis, varying $z$ from 0 to $l_{ch}$. Below cut-off ($f < 400\,\text{THz}$), spontaneous emission is completely quenched by the environment. This is due to the very low local density of states at this frequency, since there are no propagating modes in the channel. Above cut-off, we observe significant enhancement of spontaneous emission rate at the Fabry-Perot (FP) frequencies, for which the Purcell enhancement is strongly dependent on the position of the emitter inside the channel, since the density of states is localized at the nodes of the modal distribution at resonance. At the ENZ frequency $f = 400\,\text{THz}$, on the contrary, the spontaneous emission rate is not only boosted more than 100 times, but also completely uniform along the channel, as there is no resonant modal distribution at this frequency. This is qualitatively similar to the LDOS distribution experimentally probed using cathodoluminescence in [22].

In Fig. 2b we show the transverse variation of $\gamma/\gamma_0$ at the supercoupling frequency $f_{ENZ}$ at the arbitrary value $z = 0.37\,l_{ch}$. In the transverse cross-section, the LDOS is consistent with the modal distribution in the plasmonic channel, with no dependence on $y$ and a $\sim \cos(\pi x/a_{ch})$ distribution in $x$. The total spontaneous emission rate is increased by two orders of magnitude at the ENZ frequency, completely uniform in the $(zy)$ plane. This exotic feature suggests interesting possibilities when applied to a more complex quantum system, such as the possibility of enhancing the spontaneous emission of a large number of randomly located emitters, since precise positioning inside the channel would not be required. The plasmonic channel could be filled with a semiconductor matrix containing a large number of quantum dots located around the $(zy)$ plane, realizable using conventional nanofabrication techniques.

In the case of several quantum emitters located in the channel, we need to consider the complete quantum interaction in the system, which, as discussed above, is expected to be largely modified compared to a simple semi-classical model, in particular near the ENZ operation. To quantify the strength and spatial range of the quantum phenomena, we consider first the spontaneous emission of a pair of two-level quantum emitters located at two random positions along the axis of the plasmonic waveguide of Fig.1, and emitting at the ENZ frequency. We compare this situation to the case of emission at the same frequency in unbounded vacuum. The average number of decay products is given by the sum of the contributions from the symmetrical and anti-symmetrical channels of the combined system $n(t) = n_s(t) + n_a(t)$ [29]:

$$n_s(t) = (1 + \frac{\tilde{\gamma}}{\gamma})\left[1 - (1 - \frac{\tilde{\gamma}}{\gamma})^{-1} e^{-2\gamma t}(e^{(\gamma - \tilde{\gamma})t} - \frac{\tilde{\gamma}}{\gamma})\right], \quad (1)$$

$$n_a(t) = (1 - \frac{\tilde{\gamma}}{\gamma})\left[1 - (1 + \frac{\tilde{\gamma}}{\gamma})^{-1} e^{-2\gamma t}(e^{(\gamma + \tilde{\gamma})t} + \frac{\tilde{\gamma}}{\gamma})\right], \quad (2)$$

where $\gamma$ is the individual decay rate. The collective radiation damping constant $\tilde{\gamma}$ may be then calculated as a function of the separating distance $r$ and of the wave number $\beta$:

$$\tilde{\gamma} = \frac{3\gamma}{2}\left[\frac{\sin(\beta r)}{\beta r} + \frac{\cos(\beta r)}{(\beta r)^2} - \frac{\sin(\beta r)}{(\beta r)^3}\right]. \quad (3)$$

For larger separation distances $\beta r \gg 1$, $\tilde{\gamma} \to 0$ and we obtain the decay law for independent quantum emitters $n_0(t) = 2(1 - e^{-\gamma t})$. At small separations, on the other hand, $\beta r \ll 1$, $\tilde{\gamma} \to \gamma$ and we obtain Dicke's result $n_D(t) = 2\left[1 - e^{-2\gamma t}(1 + \gamma t)\right]$. Figure 3 shows the effect of distance on super-radiance for a pair of emitters placed in vacuum compared to the same pair placed in the ENZ channel, using the value of $\beta$ previously calculated. In particular, we compute $n(t_0)$, i.e., the average number of emitted photons after time $t_0 = \gamma^{-1}$, normalized to its value in the case of independent sources. The blue curve in Fig 3 refers to vacuum, while the red curve shows the calculated emission in the channel at $f_{ENZ} = 400\,\text{THz}$. For $r \to 0$ and $r \to \infty$ both curves tend to the same limits, respectively the Dicke super-radiance limit for two sources and the independent emission limit. In the channel, however, the curve drops much more slowly. At $r = 3\,\mu\text{m}$, corresponding to about four free-space wavelengths, the collective effect has totally disappeared in vacuum, while it is still evident in the ENZ channel. This is obviously associated with the large reduction in *electrical* size of the system, due to the large increase in effective wavelength when operating near cut-off. By varying the operating frequency $f_E$, the corresponding super-radiance curve changes, with its slowest drop-off expected around the ENZ frequency. In order to capture this variation, we define $L_C$ as the distance over which the collective emission stays as high as 50% of its maximum (Dicke) value. Obviously $0 < L_C < l_{ch}$, as the emitters cannot be farther away than the total available channel length. We show its variation with frequency in the inset of Fig.3 for a channel of length $l_{ch} = 1\,\mu\text{m}$. This effective Dicke length $L_C$ peaks around the ENZ frequency (it would be ideally infinite, in the case of an infinite channel and zero loss, at the cut-off frequency), and it reaches the maximum value $L_C = l_{ch}$ in a small frequency window around the ENZ operation. In this frequency range, we expect collective effects to be strong and weakly dependent on the separation distance $r$. The plasmonic channel drastically enhances the spatial range of cooperative effects, in addition to boosting the LDOS, providing an efficient way to exploit cumulative Dicke and Purcell effects over large volumes for emission frequencies sufficiently close to cut-off. The ENZ operation essentially overcomes the limitations of both effects in terms of spatial extent, and efficiently combines them to maximize their effect.

Consider finally the case of a larger number of emitters $N > 2$, randomly positioned inside the channel. At $t = 0$ this system is assumed to be in its fully excited state. If the atoms were radiating independently, the average number of photons emitted per second at time $t$ would simply be the sum of the independent contributions from all emitters:

$$I_0(t) = \frac{dn_0}{dt} = \sum_{i=1}^{N} \gamma_i e^{-\gamma_i t} \qquad (4)$$

where $\gamma_i = \gamma(\vec{r}_i)$ is the spontaneous emission rate of emitter $i$ at its position $\vec{r}_i$. Equation (4) only takes into account Purcell effects, and needs to be corrected to include collective quantum effects. Near the ENZ operation, the interactions may be taken into account following Dicke's treatment of super-radiance, whose use is justified since the system is electrically much smaller than the emission wavelength. The average intensity in the thermodynamic limit is given by the coherent hyperbolic secant pulse [31]

$$\left\langle \frac{dn}{dt} \right\rangle = \gamma \frac{N^2}{4} sech^2 \left[ \frac{\gamma N}{2}(t - t_D) \right] \qquad (5)$$

where the delay time $t_D$ depends on the total number of atoms $N$ as $t_D = \ln N / \gamma N$. At its maximum, the intensity peak reaches the value $\gamma N^2 / 4$, which, for large $N$, becomes significantly larger than the classical value $N\gamma$ obtained for a non-interacting system.

Dicke's formulation is applicable only over the narrow frequency window in which $L_C = l_{ch}$ and all emitters in the channel are involved in the collaborative effect independent of their position. At other frequencies, with $L_C < l_{ch}$, the system may be with good approximation interpreted as an array of $n \approx l_{ch}/L_C$ independent super-radiating segments, containing on average $NL_C/l_{ch}$ quantum emitters. The maximum intensity would then be in the order of

$$\langle I \rangle_{max} = \frac{l_{ch}}{L_C} \frac{\gamma}{4} \left( N \frac{L_C}{l_{ch}} \right)^2. \tag{6}$$

In this case the total radiation is no longer completely coherent, but the rate of photonic emission would still be increased by a factor equal to the cooperative gain

$$G(f) = \frac{1}{4} \frac{L_C(f)}{l_{ch}} N. \tag{7}$$

In Fig. 4 we plot the calculated maximum output intensity for a plasmonic waveguide containing an average concentration of 20 emitters per 100 nm, located in the *y-z* plane, varying the frequency and the overall length. Since the concentration of emitters is assumed constant in average, by varying the channel length we effectively change also the total number of emitters in the system. The intensity is normalized for each length to the value of emission in vacuum for the same number of emitters, so that Purcell and Dicke effects are highlighted. For frequencies around $f_{ENZ} \approx 400\,\text{THz}$ we observe a strong enhancement of the radiated intensity independently of the channel's length, with a gain reaching over 5000 for $l_{ch} = 1\,\mu\text{m}$ and $N = 200$. This gain is associated with the combination of Purcell enhancement, equal to a factor of 100 at the ENZ frequency (Fig. 2) and Dicke's quantum collective effect, contributing an additional gain $N/4 = 50$. Around ENZ, the ensemble of atoms essentially emits like a single quantum system, with a fully coherent output, despite the fact that the channel is longer than a free-space wavelength. At conventional Fabry-Perot resonances, for which Purcell effects are still quite large, the output intensity is 2 to 3 orders of magnitude smaller than its value at ENZ, due to much more localized Purcell effects and the significant shrinking of the super-radiance length. In addition, obviously these conventional resonant effects are strongly dependent on the channel length, in sharp contrast with the ENZ case, as evident in the figure. As expected from the previous analysis, the emission gain rapidly increases with the total number of emitters in the system, a phenomenon that cannot be explained in purely classical terms, highlighting the importance of cooperative quantum effects to the overall radiation enhancement, especially around the ENZ operation. It is evident that for longer channels losses may also become an important factor, quenching the overall Purcell effect and reducing $L_c$, but these results confirm that realistic losses in silver may be fully compatible with these phenomena.

To conclude, we have shown that spontaneous emission from collections of quantum sources may be drastically enhanced using narrow plasmonic channels at cut-off, due to the combination of Purcell and Dicke effects, both strongly boosted by the anomalous physics in ENZ channels, which are characterized by large enhancements of the LDOS combined with a stretched emission wavelength. The spatial extent of both Purcell and Dicke effects is therefore significantly boosted compared to emission problems in conventional materials. The overall spontaneous emission becomes totally independent of the position of the emitters within the ENZ channel, unlike other approaches to increase the local density of states based on coupling with external, localized resonances. In the setup analyzed here the collective, strongly enhanced emission from the ENZ channel is already routed in a thicker optical waveguide, and may be directed at will for further optical processing, but these concepts would work equally well if the plasmonic ENZ channel were open in free-space, as for instance in the structure proposed in [22].The proposed concept may have significant applications in various fields of research: one may amplify the weak molecular fluorescence of small samples placed inside the channel, or fill it with doped semiconductors for amplified coherent emission and low-threshold lasing in solid-state optoelectronic devices. This work has been supported by the AFOSR YIP award No. FA9550-11-1-0009 and the DTRA YIP award No. HDTRA1-12-1-0022.

1. E.M. Purcell, H.C. Torrey, and R.V. Pound, *Phys. Rev* **69**, 37 (1946).
2. S.J. Smith and E.M. Purcell, *Phys. Rev.* **92**, 1069 (1953).
3. S. Bradbury and P. Evennett, in *Contrast Techniques in Light Microscopy* (BIOS Scientific Publishers, Ltd., Oxford, U.K., 1996).
4. A.C. Pease, D. Solas, E.J. Sullivan, M.T. Cronin, C.P. Holmes, and S.P.A. Fodor, *Proc. Natl. Acad. Sci. U.S.A.* **91**, 5022 (1994).
5. V.P. Drachev et al., *J. Mod. Opt.* 49, 645 (2002).
6. S. Weiss, *Science* **283** 1676 (1999).
7. S. Nie and S. R. Emory, Science **275**, 1102 (1997).
8. A. Neogi, H. Markoc, T. Kuroda, and A. Tackeuchi, *Opt. Lett.* **30**, 93 (2005).
9. M. A. Noginov et al., *Phys. Rev. B* **74**, 184203 (2006).
10. P. Anger, P. Bharadwaj, and L. Novotny, *Phys. Rev. Lett.* **96**, 113002 (2006).
11. F. Tam, G. P. Goodrich, B. R. Johnson, and N. J. Halas, *Nano Lett.* **7**, 496 (2007).
12. R.H. Dicke, *Phys. Rev.* **93**,1, 99 (1954).
13. F.T. Arecchi and E. Courtens, *Phys. Rev. A* **2**, 5, 1730 (1970).
14. R. Fleury and A. Alù, Appl. Phys. A **109**, 781 (2012).

15. M. G. Silveirinha and N. Engheta, *Phys. Rev. Lett.* **97**,157403 (2006).

16. M. G. Silveirinha and N. Engheta, *Phys. Rev. B* **75**,075119 (2007).

17. B. Edwards, A. Alù, M. Young, M. Silveirinha, and N. Engheta, *Phys. Rev. Lett.* **100**, 033903 (2008).

18. A. Alù and N. Engheta, *Phys. Rev. B* **78**, 045102 (2008).

19. A. Alù, M. G. Silveirinha, and N. Engheta, *Phys. Rev. E* **78**, 016604 (2008).

20. A. Alù and N. Engheta, *Phys. Rev. B* **78**, 035440 (2008).

21. R. Liu et al., *Phys. Rev. Lett.* **100**, 023903 (2008).

22. E. J. R. Vesseur, T. Coenen, H. Caglayan, N. Engheta, and A. Polman, Phys. Rev. Lett. **110**, 013902 (2013).

23. C. Argyropoulos, P.-Y. Chen, G. D'Aguanno, N. Engheta, and A. Alù, Phys. Rev. B **85**, 045129 (2012).

24. A. Alù and N. Engheta, *Phys. Rev. Lett.* **103**, 043902 (2009).

25. A. Alù and N. Engheta, Materials **4**, 141 (2011).

26. P. B. Johnson and R. W. Christy, Phys. Rev. B **6**, 4370 (1972).

27. L. Novotny, B. Hecht, *Principles of nano-optics* (Cambridge University Press, New York, 2006)

28. A. F. Koenderink, Optics Letters **35**, 4208 (2010).

29. D.F. Smirnov, I.V. Sokolov and E.D. Trifonov, *Sov. Phys. JETP* **36**, 6 (1973).

30. R. Loudon, *The Quantum Theory of Light* (Clarendon, Oxford, 1983).

31. M.G. Benedict, A.M. Ermolaev, V.A. Malyshev, I.V. Sokolov and E.D. Trifonov, *Super-radiance, Multi-atomic coherent emission* (IOP Publishing, Ltd., Bristol and Philadelphia, 1996).

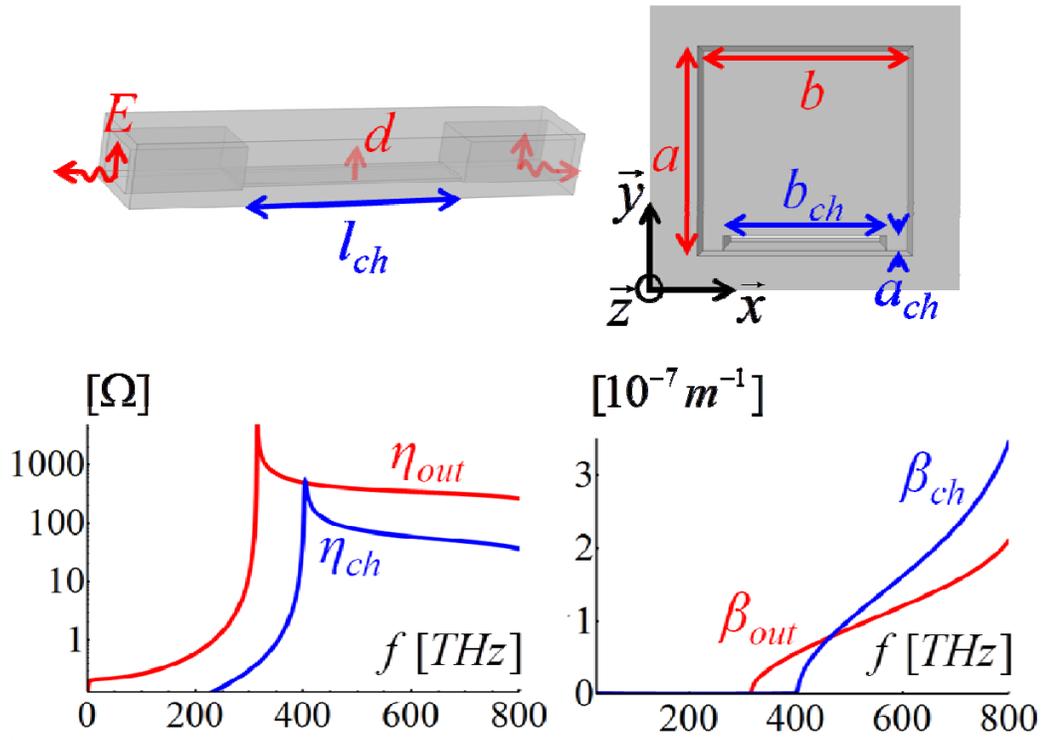

Figure 1 – Geometry of the structure: a narrow plasmonic channel connects two infinite rectangular waveguides ($a_{ch} \ll a$). The plots show the transmission-line secondary parameters, line impedance (left) and guided wave number (right).

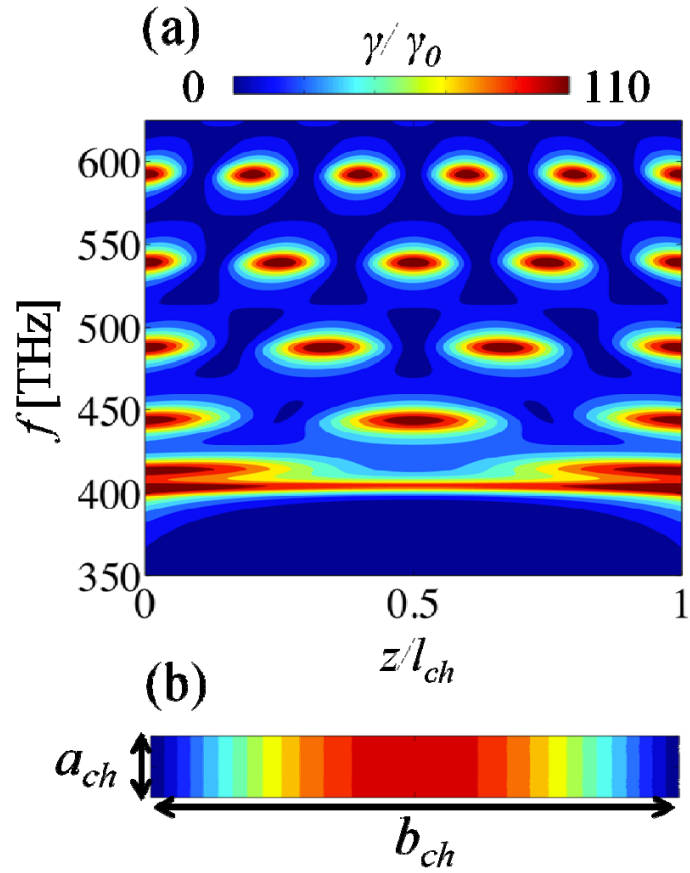

Figure 2 – Purcell effect in a plasmonic narrow channel. (a) Spontaneous emission rate versus frequency and position $z$ along the channel for a quantum emitter located on the channel axis. (b) Spontaneous emission rate in the transverse cross section of the channel at the ENZ frequency, independent of $z$.

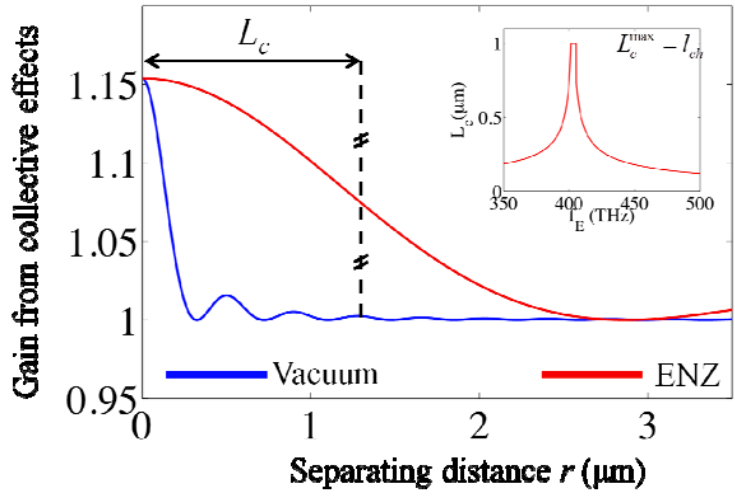

Figure 3 –Emission enhancement due to cooperative effects as a function of the separation distance between two quantum emitters in vacuum (blue curve) and embedded within the plasmonic channel of Fig. 1 at the ENZ operation (red curve). Near cut-off, the cooperation length $L_C$ is strongly increased and reaches the total channel's length $l_{ch}$.

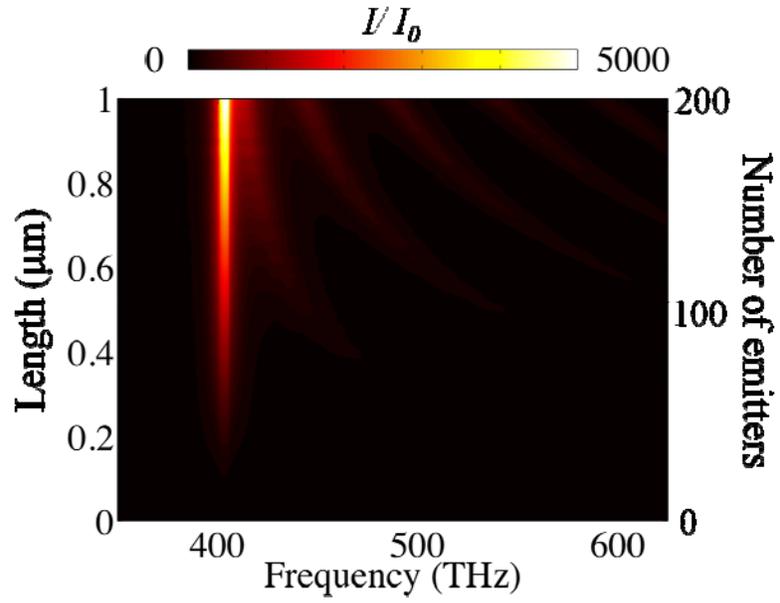

Figure 4 – Maximum emission gain compared to the same number of molecules in free-space, for an ensemble of quantum emitters randomly placed in the yz plane of the plasmonic waveguide of Fig.1, as a function of their emission frequency and the total channel length $l_{ch}$. The concentration of emitters is constant at $0.2 nm^{-1}$.